\documentclass[preprint,pra,superscriptaddress]{revtex4-1}

\usepackage[dvips]{graphicx} 
\usepackage{amsfonts}
\usepackage{amssymb}
\usepackage{amscd}
\usepackage{amsmath}    
\usepackage{enumerate}
\usepackage{epsfig}
\usepackage{subfigure}

\begin{document}

\title{Experimental measurement-device-independent quantum key distribution}


\author{Yang Liu}
\email{The authors contributed equally to the paper.}
\author{Teng-Yun Chen}
\email{The authors contributed equally to the paper.}
\author{Liu-Jun Wang}
\author{Hao Liang}
\author{Guo-Liang Shentu}
\author{Jian Wang}
\author{Ke Cui}
\author{Hua-Lei Yin}
\author{Nai-Le Liu}
\author{Li Li}
\affiliation{Shanghai Branch, Hefei National Laboratory for Physical Sciences at Microscale and Department of Modern Physics, University of Science and Technology
of China, Hefei, Anhui 230026, P.~R.~China}
\author{Xiongfeng Ma}
\email{xma@tsinghua.edu.cn}
\affiliation{Center for Quantum Information, Institute for Interdisciplinary Information Sciences, Tsinghua University, Beijing, P.~R.~China}
\author{Jason S.~Pelc}
\author{M.~M.~Fejer}
\affiliation{E.~L.~Ginzton Laboratory, Stanford University, 348 Via Pueblo Mall, Stanford CA 94305, USA}
\author{Qiang Zhang}
\email{qiangzh@ustc.edu.cn} 
\affiliation{Shanghai Branch, Hefei National Laboratory for Physical Sciences at Microscale and Department of Modern Physics, University of Science and Technology
of China, Hefei, Anhui 230026, P.~R.~China}
\author{Jian-Wei Pan}
\email{pan@ustc.edu.cn}
\affiliation{Shanghai Branch, Hefei National Laboratory for Physical Sciences at Microscale and Department of Modern Physics, University of Science and Technology
of China, Hefei, Anhui 230026, P.~R.~China}

\begin{abstract}
Throughout history, every advance in encryption has been defeated by advances in hacking with severe consequences. Quantum cryptography \cite{BB_84} holds the promise to end this battle by offering unconditional security \cite{Mayers_01,LoChauQKD_99,ShorPreskill_00} when ideal single-photon sources and detectors are employed. Unfortunately, ideal devices never exist in practice and device imperfections have become the targets of various attacks \cite{MAS_Eff_06,Qi:TimeShift:2007,Lydersen:Hacking:2010,Jain:AttackExp:2011}. By developing up-conversion single-photon detectors with high efficiency and low noise, we build up a measurement-device-independent quantum key distribution (MDI-QKD) system \cite{Lo:MIQKD:2012}, which is immune to all hacking strategies on detection. Meanwhile, we employ the decoy-state method \cite{Hwang:Decoy:2003,Lo:Decoy:2005,Wang:Decoy:2005} to defeat attacks on non-ideal source. By closing the loopholes in both source and detection, our practical system, which generates more than $25$ kbit secure key over a $50$-km fiber link, provides an ultimate solution for communication security.
\end{abstract}

\maketitle

The gap between ideal devices and realistic setups has been the root of various security loopholes  \cite{GLLP_04,BML_Squash_08}, which have become the targets of many attacks \cite{MAS_Eff_06,Qi:TimeShift:2007,Lydersen:Hacking:2010,Jain:AttackExp:2011}. Tremendous efforts have been made towards loophole-free QKD with practical devices \cite{MayersYao_98,AGM_Bell_06}. However, the question of whether security loopholes will ever be exhausted and closed still remains.

Here, we report a QKD experiment, that closes the loopholes in both source and detection and hence can achieve unconditionally secure communication. On one hand, ideal single photon sources are replaced with weak coherent states by varying mean photon intensities --- the decoy-state method  \cite{Hwang:Decoy:2003,Lo:Decoy:2005,Wang:Decoy:2005}. On the other hand, by implementing the MDI-QKD protocol  \cite{Lo:MIQKD:2012}, all the detection side channels are removed from our system.

\begin{figure}[tbh]
\centering
\resizebox{12cm}{!}{\includegraphics{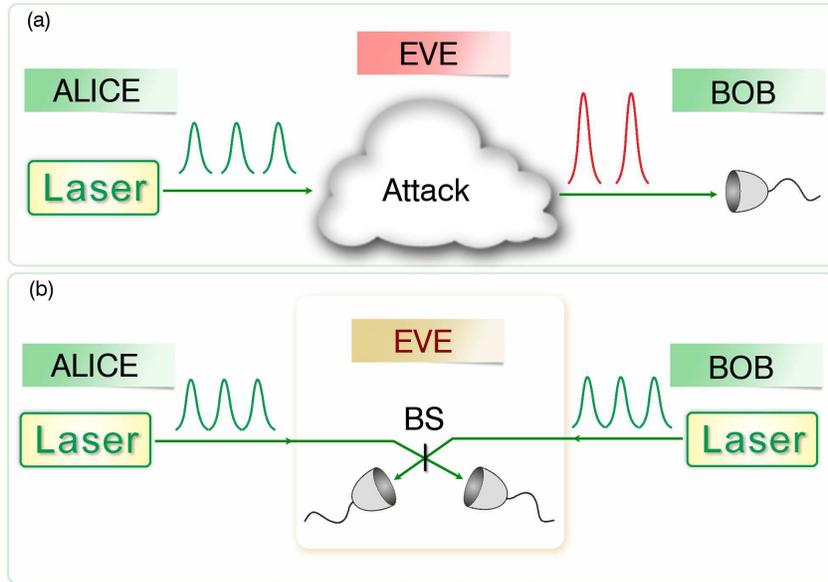}}
\caption{(a) Conventional prepare-and-measure QKD setup, where Alice sends qubits to Bob through an insecure quantum channel, controlled by Eve. (b) MDI-QKD setup, where Alice and Bob each sends quantum signals to Eve for measurement.}
\label{Fig:1}
\end{figure}

In a conventional QKD system, such as prepare-and-measure protocols, the sender, Alice, sends quantum states encoded with key information (qubits) to the receiver, Bob, who then measures them, as shown in Fig~\ref{Fig:1}a. A malicious eavesdropper, Eve, may intercept and manipulate the quantum signals traveling in the channel, and forward tampered signals to Bob. In a typical security proof of QKD \cite{GLLP_04}, one assumes that Eve performs manipulation on the Hilbert space of qubits. Since the photons have degrees of freedom other than the one used for key information encoding, Eve might take advantage of the side-channel information. For example, when an efficiency mismatch exists between detectors \cite{MAS_Eff_06}, Eve can steal some information of the key by shifting the arrival times of the quantum signals at Bob, which is called time-shift attack \cite{Qi:TimeShift:2007}.
More attacks can be launched if other degrees of freedom are considered: for instance, the detector blinding attack \cite{Lydersen:Hacking:2010,Jain:AttackExp:2011} exploits the detector's after-gate pulses and dead time.

MDI-QKD \cite{Lo:MIQKD:2012,MXF:MIQKD:2012} protocols close all loopholes on detection at once. In fact, the detectors in a MDI-QKD setup can even be assumed to be in Eve's possession. As shown in Fig.~\ref{Fig:1}b, Alice and Bob encode the key information onto their own quantum states independently and then send them to the detection station for a Bell-state measurement (BSM). The quantum signals from two arms are interfered in a beam splitter and then detected by two detectors. Certain post-selected coincidence events are used as the raw key. As discussed in Ref.~ \cite{Lo:MIQKD:2012}, even if Eve controls the measurement site, she cannot gain any information on the final key without being noticed. The security of MDI-QKD is based on the time-reversed version of entanglement-based QKD protocols \cite{Biham:ReverseEPR:1996,Inamori:Security:2002}, which is naturally immune to any attacks on detection.

\begin{figure}[tbh]
\centering
\resizebox{15cm}{!}{\includegraphics{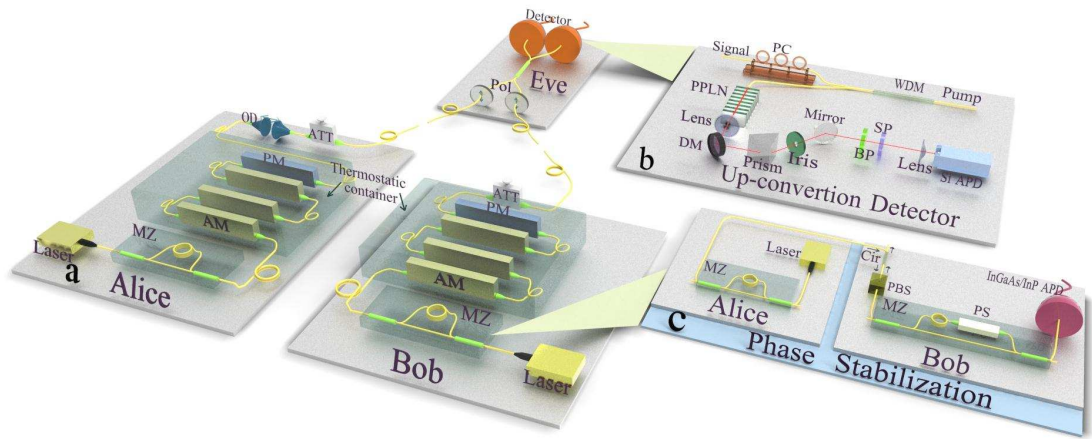}}
\caption{(a) Diagram of our MDI-QKD setup. Alice passes her laser pulses through an unbalanced Mach-Zehnder (MZ) interferometer, with an arm difference of 6 meters, to generate two time-bin pulses. A phase modulator (PM) and three amplitude modulators (AM) are used to encode the qubit and generate decoy states. All the modulations are controlled by quantum random number generators. In order to reduce the temperature fluctuation, we put all the modulators into thermostatic containers. Bob's encoding system is the same as Alice's. The pulses are then attenuated by an attenuator (ATT) and sent out via fiber links from Alice and Bob to the measurement site. After traveling through 25 km fiber spools of each arm and polarizers (Pol.), signal pulses from two sides are interfered at a 50:50 fiber beam-splitter (BS) for a partial BSM. The output photon is detected by up-conversion detectors and recorded with a time interval analyzer. (b) Diagram of an up-conversion single-photon detector. PC: polarization controller, DM: dichroic mirror, BP: band pass filter, SP: short pass filter. (c) Phase stabilization setup. Cir: circulator, PS: phase shifter, PBS: polarizing beam splitter.}
\label{Fig:Path:MIdiag}
\end{figure}

In our experimental realization, we implement the time-bin phase-encoding MDI-QKD scheme \cite{Lo:MIQKD:2012,MXF:MIQKD:2012}, as shown in Fig.~\ref{Fig:Path:MIdiag}a. Alice and Bob first randomly prepare their time-bin qubits in one of the two bases, denoted by $Z$ and $X$. If the $Z$-basis is used, the key bit is encoded in time-bin 0 or time-bin 1 by an amplitude modulator (AM). If the $X$-basis is used, the key bit is encoded into the relative phases, 0 or $\pi$, between the two time bins by a phase modulator (PM). Each party sends quantum signals to the measurement station for partial BSM. A successful BSM event occurs when the two qubits interfere perfectly in a beam splitter and the two detectors have a coincidence at alternative time bins. Then, in the $Z$-basis, a valid BSM always results in complementary bits between Alice and Bob, as is the case for the $X$-basis when each pulse contains only one photon. The multi-photon component in the coherent state pulse may cause fake coincidence, which introduces $50\%$ bit error rate in the $X$-basis. After the announcement from the measurement site, Alice and Bob will compare their basis choices and select out the sifted key (a.k.a, basis sift). Then they can perform post-processing to extract a final secure key.

A critical aspect to this experiment is the indistinguishability of the signal pulses generated by the two independent laser sources, mainly in three dimensions: spectrum, timing and polarization. Any mismatch in these dimensions would introduce errors in the $X$-basis. Firstly, the wavelength difference between Alice's and Bob's pulses need to be small compared to the bandwidth of the laser pulse. In our system, we utilize a 1 MHz shared time reference from a field-programmable gate array to modulate two independent distributed feedback (DFB) laser diodes to produce Alice's and Bob's signal pulses. The pulse width is about 2 ns and its wavelength centers at 1550.200 nm with a full width at half maximum (FWHM) of about 10 pm. By adjusting the temperature control precisely, the laser's central wavelength can be set to a precision of about 0.1 pm, which is small enough to keep the error rate low. Secondly, the temporal modes of Alice's and Bob's pulses should be overlapped precisely. We use an optical delay (OD) in Alice's station to adjust the pulse timing. The resolution of the OD is better than 10 ps and the time jitter of the laser pulses is also around 10 ps, which is small compared to the pulse width of 2 ns. Thirdly, the polarization of the quantum signals may rotate during the channel transmission due to the fiber birefringence. In front of the interference beam splitter, we insert a polarization controller and a polarizer in each arm to make the polarization indistinguishable.

The relative phase between the two arms of the Mach-Zehnder (MZ) interferometer may fluctuate with temperature and stress, which introduces further errors in the $X$-basis. We use an additional fiber between Alice and Bob as for feedback to stabilize the interferometer phases. By sending light from another laser light from Alice's MZ interferometer through Bob's MZ interferometer, we monitor the power at one of the outputs of Bob's interferometer with a single-photon InGaAs/InP avalanche photodiode (APD). The feedback is implemented by using a phase shifter inside Bob's MZ interferometer, as shown in Fig.~\ref{Fig:Path:MIdiag}c.

The performance of QKD systems is determined to a great extent by the quality of single-photon detectors, mainly in two aspects --- efficiency and noise. In comparison to the conventional QKD, MDI-QKD requires two-fold coincidence detection instead of single-fold click. Then, the channel transmittance, and hence the key rate, has quadratic dependence on the detector efficiency. Thus, high-efficiency single-photon detectors are required for MDI-QKD. Under room temperature, an up-conversion single-photon detector can provide highest quantum efficiency in telecom band. However, its dark count used to be more than 100 kHz, which limits its application in QKD. Here, we utilize long-wave pump technology \cite{Pelc:LongWave:2011} to suppress detector dark counts by two orders of magnitude. In our setup, the signal photon is mixed with a strong pump at 1940 nm in a wavelength division multiplexing (WDM) coupler and is sent to a fibre-pigtailed periodically poled lithium niobate (PPLN) waveguide, where the pump and signal interact via the sum-frequency generation process, as shown in Fig.~\ref{Fig:Path:MIdiag}b. The PPLN waveguide is a 52-mm-long reverse-proton-exchange waveguide with a poling period of 19.6 $\mu$m. A Peltier cooler based temperature-control system is used to keep the waveguide's temperature at 30 $^{o}$C to maintain the phase-matching condition.
We observe a maximum depletion of a 1550-nm input signal of 99\%, with a total internal conversion efficiency around 85\% limited by the waveguide propagation losses. The generated 850 nm photons are collected by an anti-reflection (AR)-coated objective lens, and are separated from the pump and spurious light using a combination of a short pass filter, a dichroic mirror, a prism, and a spatial filter. The light is then focused onto a commercial silicon (Si) APD with a detection efficiency of 40\% at this wavelength. Using a pump power of 200 mW, the total-system detection efficiency is 20\%, with a dark count rate of approximately 1 kHz, which can meet the stringent requirements for MDI-QKD.

At the measurement site, Eve announces the detection events when two detectors click in two different time bins. Alice and Bob post-select their key bits as the raw data according to Eve's announcement. To extract the final secure key out of the raw data, we follow the post-processing procedure presented in the Supplementary Information. When Alice and Bob respectively use average photon number $\mu$ and $\nu$, the key rate is given by the standard decoy-state formula \cite{Lo:Decoy:2005,Lo:MIQKD:2012},
\begin{equation} \label{MIExp:Post:KeyrateMI}
\begin{aligned}
R &\ge Q_{11}[1-H(e_{11})] - I_{ec}, \\
\end{aligned}
\end{equation}
where $I_{ec}$ is the cost of error correction, depending on the overall gain ($Q_{\mu\nu}$) and error rate ($E_{\mu\nu}$); $H(e)=-e\log_{2}(e)-(1-e)\log_{2}(1-e)$ is the binary Shannon entropy function; $Q_{11}$ ($e_{11}$) is the gain (phase error rate) when both sources generate single-photon states.

\begin{figure}[tbh]
\centering
\resizebox{12cm}{!}{\includegraphics{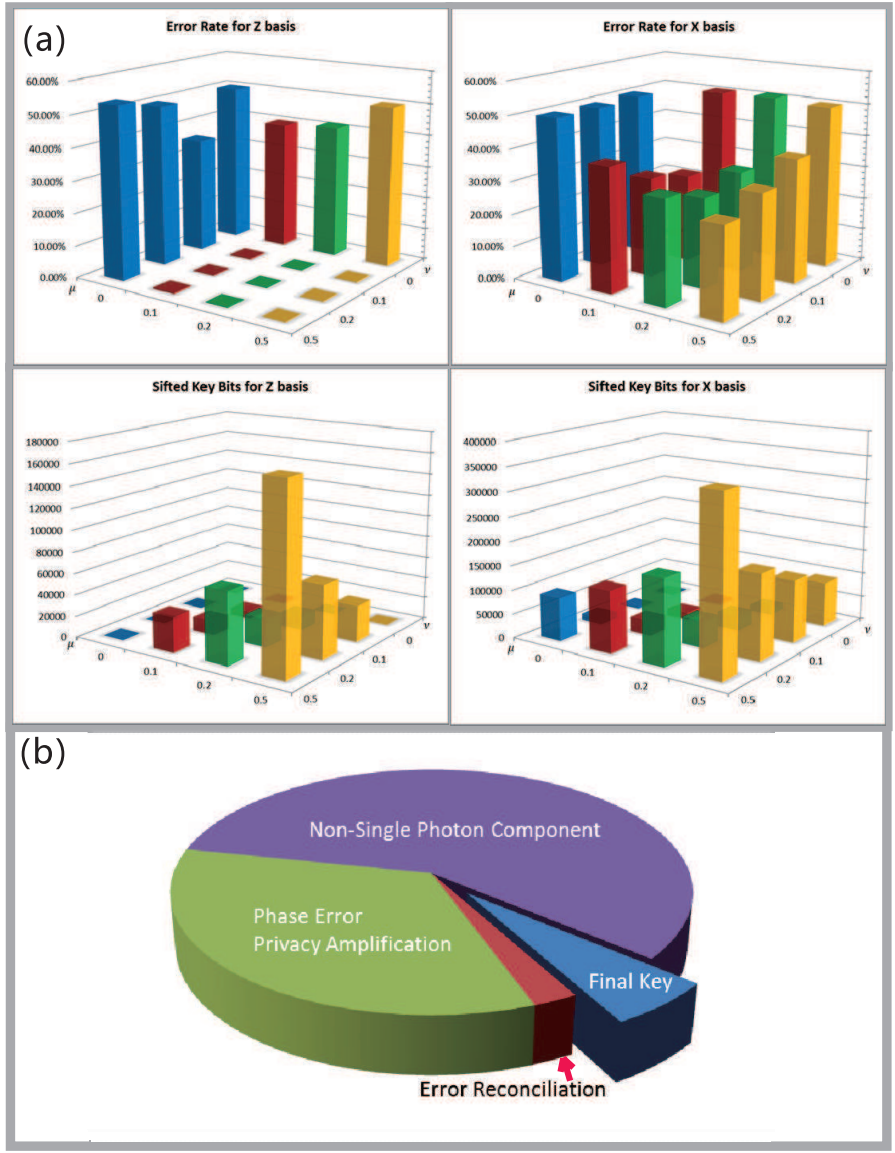}}
\caption{(a) Sifted key rate and error rate with different average photon numbers, 0, 0.1, 0.2, 0.5, in both $Z$- and $X$-basis. The data are collected by running the system for 59.5 hours. (b) Extracting secure key from the raw data. In the data post-processing, we assume 3 standard deviations for the statistical fluctuation analysis of decoy-state method. Detailed analysis is shown in the Supplementary Information.}
\label{Fig:key}
\end{figure}

In the experiment, we run our MDI-QKD system for 59.5 hours to collect raw data. Fig.~\ref{Fig:key}a shows the original sifted key bits and error rates in the $Z$-basis and the $X$-basis with different average photon numbers. From Fig.~\ref{Fig:key}a, one can see that the error rates in the $Z$-basis, $E_{\mu\nu}$, are less than 0.5\%, when the intensities, $\mu$ and $\nu$, are not 0. With the error rates (of decoy and signal states) in the $X$-basis, we can place an upper bound on the phase error rate in the $Z$-basis, i.e. $e_{11}$, which is 24.6\%\footnote{For single-photon states, the bit error probability in the $X$-basis is the same as the phase error probability in the $Z$-basis.}. Then, we evaluate the final secure key rate by Eq.~\eqref{MIExp:Post:KeyrateMI}, as shown in Fig.~\ref{Fig:key}b, from which we can see that the main reductions of the key rate come from the non-single-photon components and privacy amplification. The privacy amplification part is largely affected by the relatively small data size. Here, we have not considered the key cost in authentication and efficiency of privacy amplification, which has been shown to be small, typically, less than 1000 bits, in a practical system \cite{MXF:Finite:2011}.

\begin{figure}[tbh]
\centering
\resizebox{12cm}{!}{\includegraphics{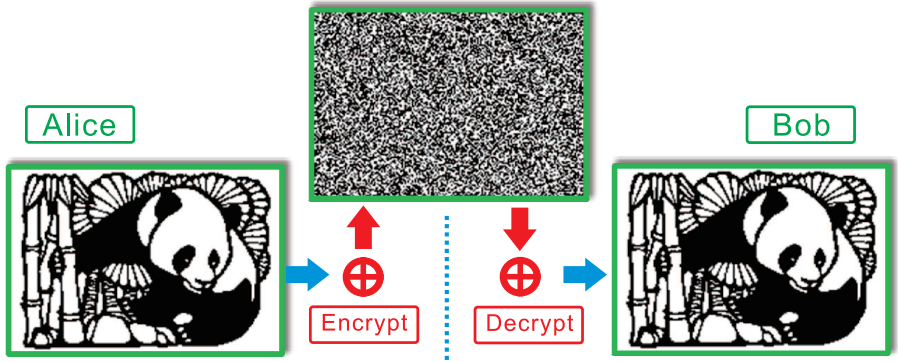}}
\caption{Demonstration of the cryptographic application of the final secure key -- one-time pad (OTP) encryption. The message, a black and white bitmap of a panda, is encrypted by the secure key generated from our system.}
\label{Fig:panda}
\end{figure}

Finally, Alice transmits a 24192-bit image to Bob via the one-time-pad protocol, using the secure key generated from our MDI-QKD system, shown in Fig.~\ref{Fig:panda}. The resulting encrypted message looks like white noise to anyone without a copy of the key, but Bob can decode it by carrying out a bitwise exclusive OR operation with his copy of the key.

We remark that the internal modulation of decoy/signal states  guarantees our system to be secure against the unambiguous-state-discrimination attack \cite{LoPreskill:NonRan:2007}. All the components in the source part, as shown in Fig.~\ref{Fig:Path:MIdiag}a, are standard commercial optical devices, which have been properly calibrated. Thus, it is reasonable to assume the side channels on source are well shielded out.

The developed up-conversion single-photon detector with high efficiency and low noise in our experiment can find immediate application in fiber based quantum technology, optical time domain reflectometer, photon-counting lidar and etc. Meanwhile, the technology of interfering two independent lasers, developed in our experiment, is also an essential building block of a quantum repeater  \cite{Zhao:Robust:2007} in global quantum communication. Furthermore, the MDI-QKD scheme can be extended into a quantum network with a star-like structure \cite{MXF:MIQKD:2012}, in which users only need photon sources but not detection systems. The expensive parts of the system, detectors, are only required at the service center, i.e., the measurement site.

The transmission distance and secure key rate can be significantly improved by increasing the repetition rate, which is mainly limited by the detector timing jitter. Our up-conversion detector can be run under a clock rate of 2 GHz \cite{Diamanti:Upconversion:2006}, with which the transmission distance can go beyond 250 km and the secure key rate can be more than 1 kbits per second at 100 km. 




\textbf{\subsection*{Acknowledgments}}
The authors would like to thank Yu-Ao Chen, Cheng-Zhi Peng, Qi-Chao Sun, Yan-Lin Tang, and Bo Zhao for enlightening discussions, especially to Chi-Hang Fred Fung for his useful discussion and comments on the key rate analysis and writing in general. This work has been supported by the National Fundamental Research Program (under Grant No. 2011CB921300 and 2011CBA00300), the NNSF of China, the CAS, and the Shandong Institute of Quantum Science \& Technology Co., Ltd.

\begin{appendix}
\section{Key rate}
In the data post-processing, we essentially follow the numerical method presented in Ref. \cite{MXF:MIFluc:2012} to extract final secure keys from the raw data. When the decoy-state protocol is used, normally, the final key is only extracted from the signal states. In our case, we extract secure keys from all decoy/signal states as long as the contribution is positive. In our experiment, we choose 4 different intensities for signal and decoy states for each arm. Thus, there are 16 combinations of detection events and the overall key rate can be given by,
\begin{equation} \label{MIExp:Key:Rsum}
\begin{aligned}
R &= \sum_{k,l=0}^{3}R_{k,l} \\
R_{k,l} &\ge \max\{Q_{11}^{\mu_k,\nu_l}[1-H(e_{11})] - I_{ec}^{\mu_k,\nu_l},0\} \\
I_{ec}^{\mu_k,\nu_l} &= Q_{\mu_k\nu_l} f H(E_{\mu_k\nu_l}),
\end{aligned}
\end{equation}
where $Q_{\mu_k\nu_l}$ ($E_{\mu_k\nu_l}$) is the overall gain (quantum bit error rate, QBER) when Alice and Bob, respectively, use expected photon numbers of $\mu_k$ and $\nu_l$; $I_{ec}^{\mu_k,\nu_l}$ is the cost of error correction with $f$ as its efficiency; $H(e)=-e\log_{2}(e)-(1-e)\log_{2}(1-e)$ is the binary Shannon entropy function; $Q_{11}^{\mu_k,\nu_l}$ is the rate for successful partial Bell-state measurement (BSM) when both sources generate single-photon states. We remark that the phase error rate of single-photon states, $e_{11}$, are assumed to be the same for all cases of signal and decoy states \cite{Lo:Decoy:2005}.

The final key is generated from the data obtained in $Z$-basis, so all the terms in Eq.~\eqref{MIExp:Key:Rsum} should be measured/inferred in $Z$-basis.  The phase error rate, $e_{11}$, in $Z$-basis, which cannot be measured directly, is inferred by the bit error rate in $X$-basis, estimated by decoy states.

\section{Parameter estimation by decoy states}
In the post-processing of MDI-QKD, the error correction term only depends on the sifted data and error correction scheme. For the privacy amplification, there are a few parameters needed to be estimated: $Q_{11}^{\mu_k,\nu_l}$ and $e_{11}$, as used in the key rate formula, Eq.~\eqref{MIExp:Key:Rsum}. Since we use weak coherent states as for quantum sources, according to the Poisson distribution of photon numbers in a coherent state, the gain of single-photon states, $Q_{11}^{\mu_k,\nu_l}$, defined as the probability that both Alice and Bob send out single-photon states with the same basis \emph{and} obtain a successful partial BSM, is given by
\begin{equation} \label{MIQKD:Model:Q11}
\begin{aligned}
Q_{11}^{\mu_k,\nu_l} = \mu_k\nu_l e^{-\mu_k-\nu_l}Y_{11}, \\
\end{aligned}
\end{equation}
where $Y_{11}$ is the yield of single-photon states, that is, the probability to get a valid BSM in the measurement site conditioned on the case when both Alice and Bob send out single-photon states. Similar to phase error rate $e_{11}$, the yield $Y_{11}$ is assumed to be the same for all signal and decoy states \cite{Lo:Decoy:2005}. The key point of the parameter estimation in the post-processing is to estimate the privacy amplification term of Eq.~\eqref{MIExp:Key:Rsum}, which depends two variables, $Y_{11}$ and $e_{11}$.

Four coherent states with different intensities are used on both Alice's and Bob's sides, $\{\mu_0,\mu_1,\mu_2,\mu_3\}$ and $\{\nu_0,\nu_1,\nu_2,\nu_3\}$, respectively. The mathematical question can be stated as follows \cite{MXFPhD},
\begin{equation} \label{MIFluc:DecoyImp:minProblem}
\begin{aligned}
\min_{\{Y_{ij},e_{ij}\}} Y_{11}^z[1-H(e_{11}^x)], \\
\end{aligned}
\end{equation}
subject to,
\begin{equation} \label{MIFluc:DecoyImp:minConstraints}
\begin{aligned}
Q_{\mu_k\nu_l}e^{\mu_k+\nu_l} &= \sum_{i,j} \frac{\mu_k^i\nu_l^j}{i!j!}Y_{ij} \\
E_{\mu_k\nu_l}Q_{\mu_k\nu_l}e^{\mu_k+\nu_l} &= \sum_{i,j} \frac{\mu_k^i\nu_l^j}{i!j!}e_{ij}Y_{ij} \\
\end{aligned}
\end{equation}
for $k, l=0,1,2,3$. There are 16 constraint linear equations, plus the constraints of $Y_{ij}, e_{ij}\in[0,1]$. We remark that the variables on the left side of Eq.~\eqref{MIFluc:DecoyImp:minConstraints} are measurable in experiment, while the right side is composed of linear functions of the unknowns, $Y_{ij}$ and $e_{ij}$. Following the numerical method presented in Ref. \cite{MXF:MIFluc:2012}, we can solve this minimization problem and estimate $Y_{11}$ and $e_{11}$ and thence the key rate given by Eq.~\eqref{MIExp:Key:Rsum}.

In order to minimize the privacy amplification term, Eq.~\eqref{MIFluc:DecoyImp:minProblem}, we calculate the lower bound of $Y_{11}$ and the upper bound of $e_{11}$, which can be solved by linear programming. Since the coefficients of $Y_{ij}$ and $e_{ij}$ decrease exponentially with growth of $i$ and $j$, this optimization problem can be solved efficiently by discarding the high order terms in the constraints of Eq.~\eqref{MIFluc:DecoyImp:minConstraints}. From our numerical evaluation, we found that the effect of the terms of $i,j\ge7$ on the parameter estimation is negligible. Apparently, it would be more efficient if we solve the minimization problem, Eq.~\eqref{MIFluc:DecoyImp:minProblem}, directly, instead of bounding $Y_{11}$ and $e_{11}$ separately. Here, we will leave it for future study.

Moreover, when the statistical fluctuations are taken into consideration, the equality constraints shown in Eq.~\eqref{MIFluc:DecoyImp:minConstraints} become inequalities. Similarly, the optimization problem can be solved by linear programming, as Eq.~\eqref{MIFluc:DecoyImp:minProblem}. In the data post-processing of our MDI-QKD system, we use 3 standard deviations for the statistical fluctuation analysis. Detailed of the analysis can be found in Ref.~\cite{MXF:MIFluc:2012}.

\end{appendix}

\bibliographystyle{apsrev}

\bibliography{Bibli}


\end{document}